\documentclass[11pt]{article}
\usepackage{latexsym}
\usepackage{mathrsfs}
\usepackage{amssymb}
\usepackage{amsmath}
\newcommand{\half}{\frac{\scriptstyle 1}{\scriptstyle 2}}
\newcommand{\A}{\mathbb{A}}
\newcommand{\C}{\mathbb{C}}
\newcommand{\CoM}{\mathbb{CM}}

\newcommand{\CP}{\mathbb{CP}}

\newcommand{\PT}{\mathbb{PT}}

\newcommand{\R}{\mathbb{R}}

\newcommand{\E}{\mathbb{E}}

\newcommand{\M}{\mathbb{M}}
\newcommand{\CM}{\mathbb{CM}}

\newcommand{\T}{\mathbb{T}}

\newcommand{\p}{\partial}
\newcommand{\dbar}{\bar\partial}
\renewcommand{\d}{\mathrm{d}}
\newcommand{\e}{\mathrm{e}}

\newcommand{\D}{\mathrm{D}}
\newcommand{\CA}{\mathcal{A}}
\newcommand{\CD}{\mathcal{D}}
\newcommand{\CF}{\mathcal{F}}

\newcommand{\cN}{\mathcal{N}}
\newcommand{\CO}{\mathcal{O}}

\newcommand{\tr}{\, \mathrm{tr}}

\newcommand{\be}{\begin{equation}\label}
\newcommand{\ee}{\end{equation}}
\newcommand{\bea}{\begin{eqnarray}\label}
\newcommand{\eea}{\end{eqnarray}}

\newtheorem{thm}{Theorem}

\begin{document}
\title{An ambitwistor Yang-Mills Lagrangian} 

\author{L.J.Mason \\ \small{ \it{The Mathematical Institute, 24-29 St
Giles, Oxford OX1 3LB, England.}\footnote{email: \tt{lmason@maths.ox.ac.uk}}}\\
D.Skinner \\ \small{ \it{Theoretical Physics, 1-3 Keble Road, Oxford
OX1 3NP, England.}\footnote{email:  \tt{skinner@thphys.ox.ac.uk}} }}

\maketitle

\abstract{We introduce a Chern-Simons Lagrangian for Yang-Mills theory
  as formulated on ambitwistor space via the Ward, Isenberg, Yasskin,
  Green, Witten construction. The Lagrangian requires the selection of
  a codimension-2 Cauchy-Riemann submanifold which is naturally picked
  out by the choice of space-time reality structure and we focus on
  the choice of Euclidean signature.  The action is shown to give rise
  to a space-time action that is equivalent to the standard one, but
  has just cubic vertices.  We identify the ambitwistor propagators
  and vertices and work out their corresponding expressions on
  space-time and momentum space. It is proposed that this formulation
  of Yang-Mills theory underlies the recursion relations of Britto,
  Cachazo, Feng and Witten and provides the generating principle for
  twistor diagrams for gauge theory.}

\section{Introduction}
Ambitwistor space $\A$ is the space of complex null geodesics in
complexified Minkowski space.  It has complex dimension five and can
be represented as the quadric $Z\cdot W=0$ inside $\PT\times\PT^*$
where $Z$ are homogeneous coordinates on projective twistor space
$\PT=\CP^3$ and $W$ are homogeneous coordinates on its dual $\PT^*$.
It has been known for many years that it is possible to reformulate
4-dimensional Yang-Mills fields onto ambitwistor space via a
generalization of the Ward transform.  A Yang-Mills connection on
space-time is encoded into a holomorphic vector $E$ bundle over some
subset of $\A$.  The Yang-Mills equations can be expressed as the
condition that the holomorphic vector bundle $E\rightarrow \A$ extends
to a certain `third formal neighbourhood' of the natural embedding of
$\A$ into $\PT\times \PT^*$, Isenberg {\it et.\ al.} (1978).  The
construction can be stated more elegantly for $\cN=3$ super-Yang-Mills
fields since the field equations hold automatically as a consequence
of integrability along super light rays.  Such super-Yang-Mills fields
correspond to holomorphic vector bundles over $\cN=3$
super-ambitwistor space $\A_{[3]}=\{Z\cdot W+\xi\cdot\eta=0 ;
([Z,\xi],[W,\eta])\in \PT_{[3]}\times\PT^*_{[3]}\}$, where
$\PT_{[3]}=\CP^{3|3}$ is $\cN=3$ super-twistor space, Witten (1978),
Harnad et.\ al.\ (1985) and Manin (1988) (and Ferber (1978) for
super-twistors).

These constructions have not, so far, been particularly useful as a
tool for studying solutions to the full Yang-Mills equations.
However, there has been recent interest arising from progress in
twistor-string theory, Witten (2004), and its spin-offs in
perturbative gauge theory, see Cachazo \& Svrcek (2005) for a review.
In addition to twistor-string theory in twistor space, a
twistor-string theory in ambitwistor space was also briefly proposed
in Witten (2004), and a number of authors, Aganagic \& Vafa (2004),
Neitzke \& Vafa (2004) and Kumar \& Policastro (2004) have argued that
there should be a mirror symmetry relation between the string theories
in twistor space and in
ambitwistor space.  Furthermore, recurrence relations for
tree-level perturbative QCD scattering amplitudes were discovered by
Britto {\it et.\ al.}\ (2005).  There it was proposed that, since the
recursion allowed one to generate arbitrary tree-level amplitudes from
trivalent ones in an ambidextrous way, the relations might be understood
as arising from a twistor-string theory in ambitwistor space.

Whilst it is expected that such a twistor-string theory should be
equivalent to a holomorphic Chern-Simons theory on (super) ambitwistor
space, it has been unclear as to how to formulate such a theory.  The
Chern-Simons form wedged against the natural super-Calabi-Yau form
only yields an $(8|6)$-form, whereas the space is $(10|6)$
dimensional.  Here we get around this problem by restricting to a
naturally defined 8-dimensional Cauchy-Riemann (CR) submanifold
$\A_\E$ of ambitwistor space $\A$ consisting of those complex null
geodesics that intersect a given real slice, here taken to be the
Euclidean slice $\E$.  Such a Cauchy-Riemann (CR) manifold has a
naturally defined analogue $\dbar_B$ of the $\dbar$-operator and
associated $\dbar_B$-Dolbeault cohomology, and the cohomology is
subject to the standard Penrose-transform with fields on space-time.
Similarly one can define CR vector bundles over $\A_\E$ and, subject
to topological triviality on the fibres, these will have a Ward
correspondence with gauge fields on $\E$.

An analytic CR vector bundle on $\A_\E$ naturally has a unique
extension to a holomorphic vector bundle on a full neighbourhood of
this submanifold in ambitwistor space, so this subspace is sufficient
to determine the full ambitwistor theory for analytic fields.
However, when the fields are non-analytic, as will generically be
the case off shell or when the field equations are satisfied in
Lorentz or split signature, there will not be any extension and the
ambitwistor theory must necessarily restrict to one on this
8-dimensional submanifold.

We give a holomorphic Chern-Simons Lagrangian for a $\dbar_B$-operator
on a bundle $E$ over the supersymmetric $\A_{[3]\E}$.  This can be
extended straightforwardly to a holomorphic Lagrangian on the CR
analogue $\PT_{[3]}\times_\E\PT^{*}_{[3]}$ of
$\PT_{[3]}\times\PT^*_{[3]}$ for $\A_{[3]\E}$.  We also give the
corresponding Lagrangians in the non-supersymmetric case and it is these
Lagrangians that we spend most time analyzing in this paper.

The Lagrangians gives a mechanism for writing down a perturbation
expansion for Yang-Mills theory involving the ambitwistor version of
the fields.  In this perturbative context, the on-shell linearised
fields can be understood as arising from standard twistor and dual
twistor cohomology classes.  We give a preliminary examination of the
ingredients of the Feynman rules: in particular we give formulae for
the propagators and vertices on ambitwistor space and their transforms
to space-time and momentum space.  We note that, being based on
Chern-Simons theory, its Feynman diagram expansion has only trivalent vertices
and this suggests that this is indeed the expansion responsible for
the BCFW recursion procedure that is generated by  trivalent vertices.
It is likely that the Feynman diagrams for this action will lead to a
generating priniciple for the twistor diagram approach to scattering
amplitudes as developed in Hodges (2005).

In \S\ref{review} we first review the standard results for
reformulating Yang-Mills theory on ambitwistor space. In
\S\ref{CRsubmanifold} we introduce the CR submanifold $\A_\E$ and
discuss the basic CR manifold theory, Ward transform and geometry.  In
\S\ref{lagrangian} we introduce the ambitwistor gauge theory
Lagrangian and explain how it fits in with the standard results.  In
\S\ref{actionequ} we give a systematic tranform of the action to
space-time and show that the action is in fact equivalent to the
following gauge theory action \be{spacetimelag} S[A,G]=\int_\E \tr(
F\wedge G -\frac\epsilon 2 G\wedge {}^*G) \ee where $A$ is a
Yang-Mills connection, $F$ its curvature, $G$ a Lie algebra-valued
2-form and $\epsilon=1/g^2$ where $g$ is the coupling constant.  The
equations of motion are $\epsilon G={}^*F$ and $DG=0$ and so it gives
rise to the standard field equations.
This gives a space-time explanation for how we can get away with a
Feynman diagram expansion using only trivalent vertices.  Note that it
is an ambidextrous analogue of the Chalmers \& Siegel (1996)
Lagrangian that bears a close relationship with twistor string theory
in twistor space, see Witten (2004), Mason (2005).  In \S\ref{pert} we
examine the perturbation theory arising from the ambitwistor
Lagrangian.  We obtain expressions for the free field inner products,
twistor transform, propagators and vertices and their transforms to
position space and momentum space.  The position and momentum space
formulae for propagators and vertices are those from the Lagrangian
(\ref{spacetimelag}).

\medskip

\noindent
{\bf Acknowledgements:} The authors would like to thank Edward Witten
for suggesting the CR submanifold $\A_\E$ and Sebastian Uhlmann for
asking an interesting question.  We would also like to acknowledge
conversations with George Sparling on CR versions of ambitwistor
theory and applications of Chern-Simons Lagrangians in twistor theory.

\section{The standard ambitwistor construction}\label{review}
An analytic Yang-Mills connection on a region in some real slice of
complex Minkowski space can be analytically continued to a connection
$A$ on a vector bundle $E'\to U$ where $U$ is a complex Stein
neighbourhood\footnote{we can and will require that the intersection
of every complex null geodesic with $U$ be connected and simply
connected.} of the given region in complex Minkowski space $\CM$.

Let $\A_U$ be the subset of ambitwistor space consisting of complex
null geodesics with non-trivial intersection with $U$.  We
construct a holomorphic vector bundle $E\to\A_U$ by defining the
fibre $E_l$ of $E$ at $l\in\A_U$ to be the vector space of covariantly
constant sections of $E'$ along the corresponding null geodesic.
We have:
\begin{thm} [Witten (1978), Isenberg, Yasskin, Green (1978)] The bundle
  $E\to\A_U$ determines and is determined by $A$.  Furthermore any
  such holomorphic vector bundle $E$ with trivial first and second
  Chern classes determines a bundle $E'\to U$ with connection $A$.
\end{thm}
Briefly, the reconstruction works by defining $E'\to U$ to be the
bundle whose fibre at $x\in U$ is the vector space of global sections
of $E$ over the corresponding quadric $Q_x$ of null directions in
$\A$.  We can define parallel propagation along  a light ray $l$ 
from $x$ to $y$ by identifying $E'_x$ with $E_l$ where $l\in Q_x$ and
then to $E'_y$ since $l\in Q_y$ also.  It turns out that this
definition of parallel propagation arises from a connection on $U$.

In order to express the field equations we need to consider the
natural embedding of ambitwistor space as a quadric hypersurface in
$\PT\times \PT^*$ where $\PT$ is projective twistor space, the
projectivisation of twistor space $\T\equiv \C^4$, and $\PT^*$ is dual
twistor space, the projectivisation of $\T^*$.  If
$(Z^\alpha,W_\beta)$, $\alpha,\beta = 0,\ldots,3$ are homogeneous
coordinates on $\PT\times\PT^*$, then $\A$ is the subset $Z^\alpha
W_\alpha=0$.  

The field equations are then expressed as follows:
\begin{thm} [Witten (1978), Isenberg, Yasskin, Green
    (1978)]\label{YMfieldeq} The 
  Yang-Mills connection $A$ satisfies the Yang-Mills equations iff the
  bundle $E$ admits an extension to a third order formal neighbourhood
  $\A_{(3)}$ of $\A$  in $\PT\times\PT^*$.
\end{thm}
There are a number of ways of saying what this means explicitly.  In
the following, this will take this to mean that there exists a smooth
bundle $E$ over $\PT\times\PT^*$ with smooth $\dbar$-operator
$\dbar_a$ satisfying $\dbar_a^2= O((Z\cdot W)^4)$ at $\A$.

There is also a formulation for super Yang-Mills.
Super-twistor space $\PT_{[3]}$ is $$\CP^{3|3}=\C^{4|3}/(Z,\xi)\sim
(\lambda Z,\lambda\xi)\, \qquad\lambda\in\C^*$$ and here $\xi^i$,
$i=1,2,3$ are the odd coordinates with $Z$ as before.  Similarly
$(W,\eta)$ are homogeneous coordinates on $\PT^*_{[3]}$ and 
$$
\A_{[3]}=
\{ ((Z,\xi),(W,\eta))\in \PT_{[3]}\times\PT_{[3]}^*| Z\cdot W+
\xi\cdot \eta=0\}.$$  We have
\begin{thm} [Witten (1978)]\label{SYMfieldeq}  The field equations for
  $\cN=3$ super 
  Yang-Mills is equivalent to the condition that the connection is
  integrable along super light-rays and this is equivalent to the
  existence of a transform to a holomorphic vector bundle $E$ over
  $\cN=3$ super-ambitwistor space $\A_{[3]}$.  Conversely, such a bundle
  determines a super Yang-Mills connection satisfying the
  integrability along super light rays, and hence the constraints and
  hence the field equations.
\end{thm}
We note that the full details of this construction are quite
complicated, see Harnad {\it et.\ al.} (1985), and for the most part
we will restrict attention to the non-supersymmetric version.

\section{The CR ambitwistor space $\A_\E$ for Euclidean
  space}\label{CRsubmanifold} Let $\E=\R^4$ be real affine Euclidean
4-space inside complex Minkowski space.  Define the 8 dimensional
submanifold $\A_\E$ of $\A$ to be the space of complex null geodesics
that intersect $\E$.  $\A_\E$ is naturally fibred over $\E$,
$p:\A_\E\to\E$ since null geodesics can only intersect $\E$ in one
point.  The fibres are the space  of complex null directions at the
point which is a complex 2-quadric $Q=\CP^1\times\CP^1$
in the projectivised complexified tangent space.  Thus,
topologically, $\A_\E=\R^4\times \CP^1\times\CP^1$.  We give an
explicit coordinatisation below.  Had we chosen a real slice of
complex Minkowski space of Lorentzian or split signature, the picture
would not be so simple as some complex geodesics intersect the real
slice in more than just one point and the corresponding points in
$\A_\M$ will generically be singular.  This is why we restrict
attention to Euclidean signature in the following.

In Euclidean signature, complex conjugation on complexified space-time
sends an $\alpha$-plane $Z$ to another, $\hat Z$, and the
complex conjugation on non-projective twistor space is in fact
quaternionic in the sense that $\hat{\hat Z}=-Z $ (i.e., complex
conjugation defines a second complex structure anti-commuting with the
standard one).  The conjugation therefore has no fixed points on the
projective space (it will be given explicitly below).  

The space $\A_\E$ is the subset $Z\cdot \hat W=0$ inside $\A$.  We
give a coordinate based derivation below.  To see this using twistor
geometry, we note that the condition $Z\cdot W=0$ is the condition
that the $\alpha$-plane corresponding to $Z$ and the $\beta$-plane
corresponding to $W$ intersect in a complex null geodesic in
complexified space-time.  The condition that $Z\cdot\hat W=0$ implies
that $Z$ actually lies on the line formed by the intersection of the
two planes $Z\cdot W=0=Z\cdot\hat W$.  But this line corresponds to
the intersection in space-time of the $\beta$-planes corresponding to
$W$ and $\hat W$ which is necessarily a point of $\E$ (or infinity) so
the complex null geodesic corresponding to $(Z,W)$ must be incident
with this real point since $Z$ and $W$ both are.

From its embedding in $\A$, $\A_\E$ inherits a CR-structure, i.e., it
has an inherited complex 3-dimensional integrable distribution $\CD$
of $(0,1)$-vectors that are the $(0,1)$-vectors on $\A$ whose real and
imaginary parts are tangent to $A_\E$.  On a CR manifold, there is a
standard construction of Dolbeault
cohomology as follows.  First define the space of
$(1,0)$-forms $\Omega^{(1,0)}$ to be the complex 1-forms that
annihilate $\CD$ and we define the $(0,p)$-forms,
$$
\Omega^{(0,p)}_B=\Omega^p/\{\Omega^{(1,0)}\wedge\Omega^{p-1}\}
$$ where $\Omega^p$ are the complex $p$-forms on $\A_\E$.  The
subscript $B$ in these definitions stands for {\em boundary} as most
studies of CR manifolds arise in the context of a real codimension-1
boundary of a complex domain.  In our situation, however, $\A_\E$ has
real codimension-2.  Note the asymmetry in the definitions,
$\Omega^{(1,0)}$ is 5 complex dimensional and is the restriction of
$\Omega^{(1,0)}$ from $\A$, whereas $\Omega^{(0,1)}_B$ is 3 complex
dimensional and defined as a quotient.  Define the $\dbar_B$ operator
to be the restriction of the exterior derivative $\d$ to
$\Omega^{(0,p)}$; $\dbar_B^2=0$ from the integrability of $\CD$.  Thus
we can define $\dbar_B$-cohomology
$$ 
H_B^p(\A_\E)=\{\mathrm{Ker}\;\dbar_B/\mathrm{Im}
\; \dbar _B \}\cap\Omega^{(0,p)}\, .
$$
We can also consider the analogue of holomorphic vector bundles
which will be a complex vector bundle $E$ with a $\dbar$-operator
$\dbar_a=\dbar_B+a$ where $a$ is a $(0,1)$-form with values in the
endomorphisms of $E$ and such that $(\dbar_a)^2=0$.  The simplest
bundles on $\A_\E$ are the line bundles $\CO(p,q)$ which are the
restrictions of tensor product of the pullback of $\CO(p)$ from $\PT$
and $\CO(q)$ from $\PT^*$ to $\A$.  The Penrose-Ward transform can be
applied in the usual way to give a correspondence between cohomology
classes or bundles on $\A_\E$ and fields or bundles with connections
on $\E$ entirely analogously with the standard ambitwistor
correspondences.  

We do not here give a complete derivation of the Ward transform in
this context, but give an indication of the main argument. Given a
topologically trivial $\dbar_B$-holomorphic vector bundle over
$\A_\E$, it must be analytically trivial over each of the fibres of
$\A_\E\to\E$ since the only topologically trivial holomorphic vector
bundle over the quadric is the trivial one.  We can define $E'\to\E$
to be the bundle whose fibre at $x\in\E$ is the corresponding space of
global sections of $E\to p^{-1}(x)$.  There is just one $(0,1)$ vector
transverse to $Q$ and, by integrability of the CR-structure it varies
holomorphically over $Q$.  It naturally has a lift to act on $\E$ and
this lift must also be holomorphic and an explicit calculation shows
that a generalization of Liouville's theorem applies to show that it
must arise from a connection on $E'\to\E$ very much as in the standard
case.

If a $\dbar_B$ cohomology class is analytic, then it extends naturally
to a cohomology class on a neighbourhood of $\A_\E$ in $\A$ so
cohomology classes on such subsets of $\A$ are determined by their
restrictions to $A_\E$ and the same is true of holomorphic bundles.
This can be seen by examining the ambitwistor correspondence: bundles
or cohomomology classes on $\A_\E$ correspond to fields on the
appropriate region in $\E$.  The field is analytic iff the
corresponding cohomology class or bundle has an analytic representative.
However, if one does have such an extension from $\A_\E$ to a
neighbourhood in $\A$, the Penrose-Ward transform will give a field on
a complex thickening of $\E$ in $\CoM$ and will therefore imply that the
corresponding field on $\E$ was analytic.  Since this is not
necessarily the case (unless one is working in Euclidean signature and
field equations are satisfied) everything is defined only on
$A_\E$ in the first instance.

The above characterisations of the field equations suggest that we
will also need to consider the embedding of $\A_\E$ into a suitable
real codimension-2 subset $\PT \times^\E \PT^*$ of $\PT \times \PT^*$.
Clearly the subset $Z\cdot\hat W=0$ extends smoothly across
$\PT\times\PT^*$ and we take this to be the definition of
$\PT\times^\E\PT^*$.  Note that the equation $Z\cdot \hat W=0$
constitutes two real non-holomorphic conditions and so it defines a CR
manifold of real codimension-2 type in $\PT\times\PT^*$.  With this
definition, $A_\E$ is the subset of $\PT \times^\E \PT^*$ on which
$Z\cdot W=0$.

\subsection{The fibrations over $\E$ and explicit
  coordinatization}\label{coords} 
Here we develop further the geometry of the embedding of $\A_\E$ in
$\PT\times^\E\PT^*$ in the context of the projections of $\PT$,
$\PT^*$ and $\A_\E$ to euclidean space $\E$ and its conformal
one point compactification $S^4$.  

In euclidean signature, twistor space and its dual have projections to
$S^4$.  The most primitive definition of a (dual) twistor is as a
totally null (anti) self-dual 2-plane or $\alpha$ ($\beta$)-planes in
complex Minkowski space.  These intersect $S^4$ at precisely one point
and this leads to fibrations $p:\PT\to S^4$, $p:\PT^*\to S^4$ where we
have abused notation to call all such fibrations $p$.  The fibres of
these fibrations are $\CP^1$s and can naturally be identified with
projective (anti) self-dual spinors.  The non-projective twistor space
$\T-\{0\}$ can be identified with the total space of the bundle of
self-dual spinors (minus the zero-section) and similarly $\T^*-\{0\}$
can be identified with the complement of the zero-section in the
bundle of anti-self-dual spinors.  In Atiyah, Hitchin \& Singer
(1978), twistor spaces were defined in a similar way as the the total
spaces of bundles of metric and orientation compatible almost complex
structures.  There they were represented as self-dual 2-forms of unit
length, whereas here we use the representation of a metric compatible
almost complex structure by a spinor.  See Woodhouse (1985) for an
introduction to the approach used here and a review of basic twistor
theory in this context.

We introduce coordinates $(x^a,y^b), a,b=0,1,2,3$ on $\E\times\E$ and
 coordinates
 $(Z^\alpha,W_\beta)=((\omega^A,\pi_{A'}),(\lambda_B,\mu^{B'}))$ on
 $\PT\times \PT^*$, $A=0,1$ and $A'=0',1'$. We have $Z^\alpha W_\alpha
 = \omega^A\lambda_A+\pi_{A'}\mu^{A'}$.  The incidence relations with
 spacetime are
$$
\omega^A=x^{AA'}\pi_{A'}\, , \qquad \mu^{A'}=-y^{AA'}\lambda_A\, ,
$$
where $x^{AA'}=\sigma_b^{AA'}x^b$ and $\sigma_b^{AA'}$ are the
standard Van de Waerden symbols.

The Euclidean complex conjugation induces the spinor conjugation
$\omega^A\rightarrow \hat\omega^A=(\bar \omega^1, -\bar\omega^0)$ 
and $\pi_{A'}\rightarrow \hat\pi_{A'}=(\bar \pi_{1'},
-\bar\pi_{0'})$.  This extends to the conjugations
$$
Z\rightarrow \hat Z= (\hat \omega^A,\hat\pi_{A'})\, , \qquad
W\rightarrow \hat W=(\hat\mu^{A'},\hat\lambda_A)
$$
With this notation, the fibration $p:\PT \times \PT^*
\rightarrow\E\times\E$ is given by
$$
p(Z^\alpha,W_\beta)=\left( \frac1{\hat\pi^{B'}\pi_{B'}}(\omega^A\hat
\pi^{A'}- \hat\omega^A\pi^{A'}),
\frac{-1}{\hat\lambda^B\lambda_B}(\mu^{A'}\hat\lambda^A
  -\hat\mu^{A'}\lambda^A)\right) \, .
$$
The functions $Z\cdot W=(x^{AA'}-y^{AA'})\pi_{A'}\lambda_A$, and
$Z\cdot\hat W=(x^{AA'}-y^{AA'})\pi_{A'}\hat \lambda_A$.

The fibres of $p:\PT\times\PT^*\to\E\times\E$ are the cartesian
product of the Riemann spheres parametrised by homogeneous coordinats
$\pi_{A'}$ and $\lambda_A$, so we can equivalently use the
non-holomorphic coordinates
$\left((x^{AA'},\pi_{A'}),(y^{BB'},\lambda_B)\right)$ on
$\PT\times\PT^*$.
The distribution $\CD$
defining the (0,1) vectors is given by
$$\CD=\langle\pi^{A'}\frac{\p}{\p x^{AA'}}, \lambda^A\frac{\p}{\p
  y^{AA'}}, \frac{\p}{\p\hat \pi_{A'}}, 
\frac{\p}{\p\hat\lambda^A }\rangle$$
The distribution can be equivalently defined as being those vectors
that are orthogonal to $D^3Z\wedge D^3W$ where
$$
D^3Z=\D\pi\wedge \pi_{B'}\pi_{C'}\d^2x^{B'C'}
\quad \mbox{ and }\quad
D^3W=\D\lambda \wedge \lambda_B\lambda_C\d^2y^{BC} 
$$
where we define $\D\pi= \pi_{A'}\d\pi^{A'}$, 
$\D\lambda=\lambda_A\d\lambda^A$, and
$$
\d^2x^{B'C'}=
\varepsilon_{BC}\d
x^{BB'}\wedge \d x^{CC'}\, ,\quad
\d^2y^{BC} =
\varepsilon_{B'C'}\d
y^{BB'}\wedge\d y^{CC'}.
$$

On $\A_\E$ we have $x^a=y^a$ and hence homogeneous coordinates $(x^a,
\pi_{A'},\lambda_A)$ (we leave it to the reader to check that $Z\cdot
W=Z\cdot\hat W=0$ implies that $x^a=y^a$).  The distribution defining
the $(0,1)$ vectors non-projectively is $\{\pi^{A'}\lambda^A\p/\p
x^{AA'}, \p/\p\hat \pi^{A'} , \p/\p\hat\lambda^A\}$.  The $(1,0)$
forms are spanned by $\{\pi_{A'}\d x ^{AA'}, \lambda_A\d x^{AA'},
\d\pi_{A'}, \d\lambda_A\}$.

In this notation we define 
$$
\PT\times^\E\PT^*= \{ (Z,W)\in\PT\times\PT^*| p(Z,W)\subset\E\times
\E, Z\cdot\hat W=0\}\, .
$$ We emphasize that the notation $\PT\times^\E\PT^*$ is {\em not }
intended to indicate the fibre-wise product.  Whereas $p:\A_\E\to\E$,
$p:\PT\times\PT^*\to S^4\times S^4$ induces a fibration $p:\PT
\times^\E \PT^*\to\E\times \E$.  On $\PT\times^\E\PT^*$ we have that
the fibres over $\E\times \E$ are $\CP^1\times \CP^1$ as above on
$\A_\E\to \E$ where here $\E\subset\E\times\E$ as the diagonal, but
away from the diagonal, the fibres jump to  $\CP^1$ since the
constraint $Z\cdot\hat W=0$ gives $s^{AA'}\pi_{A'}\hat\lambda_A=0$
where $2s^a=x^a-y^a$.  When $s^{AA'}\neq 0$, we can solve for
$\lambda_A$ up to scale
$$
\lambda^A\propto s^{AA'}\hat\pi_{A'}\, .
$$
Thus, projectively, we can coordinatise the fibre with just
$\pi_{A'}$. 
The $(1,0)$-forms are still spanned by $\{\pi_{A'}\d x ^{AA'},
\lambda_A\d x^{AA'}, \d\pi_{A'}, \d\lambda_A\}$.  


\section{The Lagrangian}\label{lagrangian}
Let $E\rightarrow \PT\times^\E\PT^*$ be a smooth complex vector
bundle.  The field variable is a $(0,1)$-form $a$ with values in the
endomorphisms of $E$ so as to define an extension of $\dbar_B$ to an
operator $\dbar_a$ on sections of $E$ (we drop the subscript $B$ as
all $\dbar$ operators in the following will be $\dbar_B$ operators).
We give two versions of the action, the first being \be{ambilag}
S[a]=\int_{\PT \times^\E \PT^*} \tr \left(F^{(0,2)}\wedge F^{(0,2)}
\right) \wedge \frac {D^3 Z\wedge D^3W }{(Z\cdot W)^4 } \ee where
$D^3Z=\varepsilon_{\alpha\beta\gamma\delta}Z^\alpha \d Z^\beta\wedge
\d Z^\gamma \wedge \d Z^\delta$, $D^3W$ is defined similarly and
$F^{(0,2)}=\dbar_a^{2}$.  This action is related to a holomorphic
Chern-Simons action with an integration by parts since $\dbar_B \mathrm{CS}(a)
= \tr (F^{(0,2)})^2$ where $\mathrm{CS}(a)=\half \tr(a\wedge \dbar a+
\frac23 a^3)$, so formally \be{CSlag}
S[a]=\int_{\PT\times^\E\PT} \mathrm{CS}(a) \wedge
\delta^{\prime\prime\prime}(Z\cdot W) D^3 Z D^3 W\, .  \ee where
$\delta^{\prime\prime\prime}(Z\cdot W)=\dbar \left(\frac
1{(Z\cdot W)^{4}}\right)$ is the third holomorphic derivative of the delta
function.

Here we are defining, for a complex variable $z$,
$$
\delta (z)=\dbar \left(\frac 1z\right)= 2\pi i\delta(\Re z)\delta(\Im z) \d
  \bar z\, , \quad \mbox{ and } \quad \delta'(z)= \frac \p{\p z}\delta(z)
$$ and so on.  It can be checked that the homogeneity properties of
$\delta$-functions are such that $\delta(\lambda z)=\delta(z)/\lambda$
and so on for the derivatives.  Thus, since
$Z\cdot W$ takes values in the line bundle $\CO(1,1)$, $\delta(Z\cdot
W)$ makes
sense as a $(0,1)$-form with values in $\CO(-1,-1)$.  Similarly
$\delta^{\prime\prime\prime}(Z\cdot W)$ is a $(0,1)$-form with values
in $\CO(-4,-4)$.

This second formulation leads to the ambitwistor formulation of the
Yang-Mills equations of theorem \ref{YMfieldeq} directly
in the form
$$
F^{(0,2)}\wedge\delta^{\prime\prime\prime}(Z\cdot W)=0\, ,
$$
i.e., 
$F^{(0,2)}=O(Z\cdot W^4)$, vanishing to 3rd order about $Z\cdot W=0$.

Note that, since the action integral is supported to 3rd order on
$\A_\E$, any variation $a\to a+\delta a$ with $\delta a=O(Z\cdot
W^4)$ is a gauge symmetry over and above the standard gauge symmetry.

\subsection{The supersymmetric version}

It is also interesting to consider a supersymmetric version of this action. 
We start with the $\mathcal{N}=3$ supertwistor space $\T_{[3]}=\C^{4|3}$ with
coordinates $(Z^\alpha,\xi^i)$, $i=1,2,3$ where $\xi^i$ are Grassmann-odd. 
The dual supertwistor space $\T^*_{[3]}$ likewise has coordinates 
$(W_\beta,\eta_j)$ and we projectivise with the usual equivalence 
$(Z^\alpha,\xi^i)\sim(\lambda Z^\alpha,\lambda\xi^i)$, $\lambda\in\C^*$. 
Super ambitwistor space is then defined as
$$
\A_{[3]} = \left\{([Z^\alpha,\xi^i],[W_\beta,\eta_j])\in\PT_{[3]}\times
\PT^*_{[3]}\,;\,Z^\alpha W_\alpha+\xi^i\eta_i =0\right\}.
$$
Super ambitwistor space is (for $\mathcal{N}=3$) a Calabi-Yau supermanifold
as it possesses a global holomorphic measure
$$
\Omega
=\oint\frac{D^3Z\wedge{\rm d}^3\xi\wedge D^3W\wedge{\rm d}^3\eta}{
Z\cdot W+\xi\cdot\eta},
$$
where the contour is taken around any $S^1$ encircling $\A\in\PT\times\PT^*$.
The integrand has weight $(0,0)$ under the scaling of homogeneous coordinates
(since ${\rm d}\xi\mapsto\lambda^{-1}{\rm d}\xi$ by the rules of Berezinian 
integration) and has a pole on $Z\cdot W+\xi\cdot\eta=0$. Hence it localizes 
on $\A_{[3]}$ after performing the contour integral, and defines a holomorphic
measure.

As above, we can write our action on the ambient space 
$\PT_{[3]}\times^\E\PT^*_{[3]}$
$$
S[a_s]=\int_{\PT_{[3]}\times^\E\PT^*_{[3]}}{\rm tr}(F^{(0,2)})^2
\wedge\frac{D^3Z\wedge {\rm d}^3\xi\wedge D^3W\wedge{\rm d}^3\eta}{
Z\cdot W+\xi\cdot\eta},
$$
where now $a_s\in\Omega_B^{(0,1)}\otimes{\rm End}(E)$ defines a 
$\bar\p_B$-operator on the smooth complex bundle 
$E\rightarrow\PT_{[3]}\times^\E\PT^*_{[3]}$. Integrating by parts as above, 
we obtain a holomorphic Chern-Simons theory directly on $\A_{[3]\E}$ with 
action
$$
S[a_s]=\int_{\A_{[3]\E}}\mathrm{CS}(a_s)\wedge\Omega
$$ where we have used $\bar\p(Z\cdot W+\xi\eta)^{-1}= \delta(Z\cdot
W+\xi\cdot\eta)$ and integrated out the $\delta$-function.  Thus, in
the supersymmetric theory, all the information is present on
$\A_{[3]\E}$ and we do not have to consider the ambient
$\PT_{[3]}\times^{\E}\PT^*_{[3]}$. Essentially, this is because the
Grassmann-odd directions mimic the effect of the extension to the
third formal neighbourhood, Eastwood \& LeBrun (1986). We can recover
the non-supersymmetric theory by considering only those terms in
$\mathcal{A}$ which are independent of $\xi$ and $\eta$, or
equivalently, by setting the superpartners of the standard spin-1
gauge field to zero. In this case, the only terms which survive the
Grassmann integration come from the term $(\xi\cdot\eta)^3/(Z\cdot
W)^4$ in the expansion of $(Z\cdot W+\xi\cdot\eta)^{-1}$, and we
recover our non-supersymmetric action.


\section{Relation to standard Yang-Mills action}\label{actionequ}
It is clear from the above that the solutions to the field equations
arising from the action principles given above should give rise to
solutions of the Yang-Mills equations.  In this section we
give an explicit proof and show that the actions can in fact be
identified, at least on solutions to the equations of motion.

Given a Yang-Mills connection $A$ on a bundle $E'\rightarrow \E$, we
consider an arbitrary smooth extension $\CA$ to $\E\times \E$ where
$\Delta:\E\rightarrow \E\times \E$ is embedded as the diagonal.  We
will restrict the action to $\dbar_B$ operators on bundles over
$\PT\times^\E\PT^*$ that are induced from such connections pulled back
from $\E\times\E$.  The reason we can assume that our
$\dbar_B$-operators are obtained in this way is because the part of
the field equation $\dbar_a^2=O(Z\cdot W^2)$ on $\A_\E$, together with
the gauge freedom to alter $a$ at higher order away from
$\A_\E$ will imply that the $\dbar_B$-operator is gauge equivalent to
one obtained from a connection pulled back from $\E\times \E$ in this
way, and so this will necessarily be the case when some of the
ambitwistor field equations are satisfied (although the Yang-Mills
equations themselves are not implied at this point).

The action then becomes:
$$
S[A]=\int_{\PT\times^\E\PT^*} \tr \; \CF^2 \wedge \frac{D^3Z\wedge D^3
  W}{(Z\cdot W)^4}
$$
where $\CF$ is the curvature of the connection $\CA$.  On
$Z\cdot\hat W=0$, we have that
$\lambda^A=(x^{AA'}-y^{AA'})\hat\pi_{A'}$ up to scale and, since the
integrand is scale invariant, we can eliminate $\lambda_A$ and then
integrate over the $S^2$ fibres away from the diagonal over $\E\times
\E$.  Introduce 
$$
s^a=\frac12(x^a-y^a)\, , \qquad r^a=\frac12(x^a+y^a)\, .
$$
We have
$$
\int_{s^{AA'}\lambda_A\hat\pi_{A'}=0} \D\lambda\wedge\D\pi
\frac{\pi_{A'}\pi_{B'}\lambda_A\lambda_B}{(2s^{AA'}\lambda_A\pi_{A'})
 ^4}=16\pi i\frac{s_{A(A'}s_{B')B}}{3(s\cdot s)^3}
$$
where the integral is over the $S^2$ fibres away from the
diagonal over $\E\times\E$ (this follows in particular from \S3 of
Woodhouse 1985).  Thus, performing the fibre integrals we obtain
$$
\int_{Z\cdot\hat W=0} \frac{
D^3Z\wedge
D^3 W}{(Z\cdot W)^4}=16\pi i
\frac{s_{AA'}s_{BB'}}
{(s\cdot s)^3}\d^2x^{A'B'}\d^2y^{AB}
$$  
so that  we obtain the action for the
connection $\CA$:
$$
S[\CA]=\int_{\E\times\E} \tr \CF^2 \wedge \frac
{s_{AA'}s_{BB'}\d^2x^{A'B'}\d^2y^{AB}}{(s\cdot s)^3} 
$$ on $\E\times\E$ (recall that we have assumed that the connection
itself has no dependence on the fibre coordinates).  We now choose a
frame for the bundle, write $\tr\CF^2=\d\mathrm{CS}(\CA)$ and
integrate by parts.  We have the relation
\begin{align*}
\d \left( \frac{s_{AA'}s_{BB'}\d^2x^{A'B'}\d^2y^{AB}}{(s\cdot
    s)^{-3}}\right)&=\frac{8\pi^2}3 \frac{\p
\delta^4(s)}{\p s^{AA'}}\;
    \d s_{BB'}\wedge \d^2x^{A'B'}\wedge\d^2y^{AB}
\\&= - 2\pi^2 \frac{\p
\delta^4(s)}{\p s^{AA'}}
\left(\d^3x_B^{A'}\d^2y^{AB} + 
  \d^2x^{A'B'}\d^3y^A_{B'}\right)
\end{align*}
where $\d^3x^a={}*\d x^a=\frac16\varepsilon^a{}_{bcd}\d x^b\wedge\d
x^c\wedge \d x^d$; the formula follows from the
derivative of the standard relation $\nabla_a\nabla^a(1/s\cdot
s)=4\pi^2\delta^4(s)$. 
Integrate by
parts on the $\p/\p s^{AA'}$ derivatives to obtain the action
(ignoring irrelvant overall constant factors)
\begin{align*}
S[\CA]&=\int_{\E\times\E} \frac{\p \mathrm{CS}(\CA)}{\p s^{AA'}} \wedge
\delta^4(s) \;\d s_{BB'}\wedge \d^2x^{A'B'}\wedge\d^2y^{AB}
\\
&=\int_{\E\times\E} \tr \left(\frac{\p \CA}{\p s^{AA'}} \wedge\CF
- \frac12 \d \left(\CA\wedge\frac{\p\CA}{\p s^{AA'}}\right) \right)
\delta^4(s) \; \d s_{BB'}\wedge \d^2x^{A'B'}\wedge\d^2y^{AB}
\end{align*}
We now integrate by parts on the second term to obtain an
expression in the $\p\delta^4(s)/\p s^{CC'}$ and integrate by parts
back on the $\p/\p s^{AA'}$ derivatives again
to  obtain
$$
S[\CA]=\int_{\E\times\E} \tr \left(\frac{\p \CA}{\p s^{AA'}} \wedge\CF
- \frac12\frac{\p}{\p s^c} \left(\CA\wedge\frac{\p\CA}{\p
    s^{AA'}}\right)\d s^c \right) 
\delta^4(s) \; \d s_{BB'}\wedge \d^2x^{A'B'}\wedge\d^2y^{AB}
$$
The indices in the second term turn out to be skew over $a$ and $c$
and so no second derivative terms in $s^a$ appear. We can now
integrate out the $s^a$ using the delta functions to obtain the action
as a functional of $\CA$ and its first derivative on $s^a=0$.  To
simplify the calculation, choose an arbitrary gauge at $s^a=0$ and
parallel propagate the frame out along $s^a(\p/\p x^a-\p/\p y^a)$.
Then set $\CA=A^x_a\d x^a + A^y_a\d y^a$ and expand in $s^a$
\be{expansion} 
A^x_a= \half A_a +A^{x}_{ab}s^b + \ldots \quad A^y_a=
\half A_a +A^{y}_{ab}s^b + \ldots \ee where $A_a(r^b)$ is a given
gauge field on $\E$, and $A^x_{ab}$ and $A^y_{ab}$ are functions of
$r^a$ only and the gauge condition (which has already been used in the
above) also implies that $A^x_{(ab)}-A^y_{(ab)}=0$.  The action
reduces to a functional of $A_a$, $A^x_{ab}$ and $A^y_{ab}$.
Decomposing into irreducibles, we discover that some of the
irreducible pieces of $A^x_{ab}$ and $A^y_{ab}$ decouple from the
other fields and appear quadratically in the action so that the field
equations make them vanish.  Eliminating these we are left
with $A_a$ and a Lie algebra valued 2-form $G$ with action
\be{spacetimeaction}
S[A, G]=\int \tr G\wedge F - \frac12 \int\tr \; G\wedge {}^*G\, ,
\ee
where $F$ is the curvature of the connection $A$.  The field equations
from this action are simply $G={}^*F$ and $DG=0$ and so we see that,
eliminating $G$ from the action, we finally obtain the standard
Yang-Mills action.

\section{Perturbation theory}\label{pert}
The action allows us to consider perturbation theory.  In perturbation
theory we take the $(0,1)$-form $a$ to be `on-shell' in the sense that
it satisfies the field equations of linear theory.  In linear theory
we can take $a$ to be a Lie algebra valued element of $H^1(\A_\E,\CO)$.
The extension of such a cohomology class to 3rd order in
$\PT\times^\E\PT^*$ implies extension to all orders (this can be seen
by the sheaf theoretic arguments of Baston and Mason 1987).  By the
Kunneth formula (or simply by the Penrose transform and the fact that
linear fields decompose into self-dual and anti-self-dual parts) we
can write $a=a^-(Z) + a^+(W)$ and the notation indicates that $a^-(Z)$
is pulled back from a class in $H^1(\PT',\CO)$, i.e., $a^-(Z)$ is a
$(0,1)$-form pulled back from twistor space and  $a^+(W)$ is one
pulled back from dual-twistor space.  ($\PT'=p^{-1}(\E)$ is $\PT$ with the line
corresponding to the point at $\infty$ taken away---the cohomology of
$\PT$ is at most finite dimensional.)

To construct Feynman diagrams from this action, we note that the
vertices coming from the Chern-Simons action are all trivalent.
(The possible 4-vertex coming from the $\tr(F^2)$ form of
the action vanishes identically.)

\subsection{The inner product, twistor transform and propagators}
The symplectic inner product on such linear fields can be derived from
the Lagrangian in the standard way as the boundary term in the
variation of the action.  This gives
\be{symplecticform}
\langle a_1,a_2\rangle= \int_{\mathrm{Contour}} \tr\, (a_1\wedge a_2) \;
\delta^{\prime\prime\prime}(Z\cdot W) D^3Z\wedge D^3 W
\ee
This is the integral of a closed 9-form over the contour obtained by
the intersection of Euclidean space with Lorentzian space.  Using the
delta-function to restrict the integral to $\A_\E$, the contour can be
taken to be the part of $\A_\E$ fibering over this intersection of the
Euclidean with the Lorentzian slice.  This symplectic inner product
yields a positive definite inner product on positive frequency fields
on Minkowski space: such fields extend over the $t\geq 0$ half of
Euclidean space (where $t=0$ is the intersection with Minkowski space)
and taking $a_1$ to be such a positive frequency field, and $a_2$ to
be its Minkowskian complex conjugate, so that it extends over $t\leq
0$, the above integral gives the standard norm.

In twistor theory there is an alternative representation of helicity
$\pm 1$ fields in terms of twistor functions of homogeneity degree
$-4$, $g^+(Z)\in H^1(\PT',\CO(-4))$ and $g^-(W)\in
H^1(\PT^{*\prime},\CO(-4))$.   These are related to the homogeneity
degree 0 representation by
$$
g^+(Z)=\int_{Z\cdot\hat W=0} a^+(W)\wedge
\delta^{\prime\prime\prime}(Z\cdot W) D^3W\, . 
$$ This is an integral of a 5-form over the four-manifold $W\cdot\hat
Z=0$ and can be checked explicitly using the explicit Dolbeault
representatives of Woodhouse (1985) in which $a^+(W)=
A(y)_{AA'}\lambda^A\hat\lambda_B\d y^{BA'}$ and $g^+(Z)=\hat
\p(\phi_{A'B'}(x) \hat\pi^{A'}\hat\pi^{B'}/(\hat\pi\cdot\pi)^3)$,
$\phi_{A'B'}=\nabla^A_{(A'}A_{B')A}$ and $\hat\p=\d \hat Z^\alpha\p/\p
Z^\alpha$ (here we use the notation of \S\ref{coords}).  

It is worth noting that this twistor transform relation also works in
this context off-shell, i.e., $\dbar a$ does not need to vanish
although, in that case, $\dbar g^+$ will not vanish either, but the
push down of $g^+\wedge D^3Z$ from $\PT$ to $\E$ will be the self-dual
2-form corresponding to the potential $A^+$ on $\E$ whose pullback to
$\PT^*$ and projection onto $\Omega^{0,1}$ determines $a^+(W)$.

Using the twistor transform, we can verify that the symplectic inner
product gives the standard expression by taking $a_1$ to be pulled
back from $\PT^*$ and $a_2$ from $\PT$ to give
\begin{align*}
\langle a_1(W),a_2(Z)\rangle&= \int_{t=0} \tr\,
(a_1(W)\wedge a_2(Z)) \; 
\delta^{\prime\prime\prime}(Z\cdot W) D^3Z\wedge D^3 W\\
&=\int \tr\, (g_1(Z)\wedge a_2(Z))\wedge D^3Z\, .
\end{align*}
This is a standard expression for the inner product in twistor theory
in which the $t=0$ contour becomes $PN$, the space of null twistors
in Lorentz signature that correspond to real light rays in Minkowski space.

The Chern-Simons propagator $\Delta$ is formally $\dbar^{-1}$ acting
on $\Omega^{0,1}$ with appropriate boundary conditions. If we write
$\dbar a= j$ where the current $j$ is a closed $(0,2)$-form, then we
have that currents and potentials are dual by
$$
(j,a)= \int a\wedge j\wedge \delta^{\prime\prime\prime}(Z\cdot W)
D^3Z\wedge D^3W\, .
$$
The propagator then can be expressed as the integral kernel $\Delta$
satisfying the formal relation
\be{propdefn}
 j_1 \Delta j = (j_1,a) \;\;\forall j_1 \qquad \Leftrightarrow\qquad
 \dbar a=j. 
\ee
We have the following formula for $\Delta$
$$ 
j_1\Delta j_2=\int j_1\wedge j_2 \frac {D^3Z\wedge D^3 W}{(Z\cdot W)^4}\, ,
$$
and (\ref{propdefn}) can now be verified directly by integrating by parts.

\subsection{Space-time and Momentum space Feynman rules}
It is most likely that the direct application of Feynman rules to this
action yields a version of the twistor diagram formulation of
scattering amplitudes; see Hodges (2005) for a recent discussion.
Indeed it is to be hoped that this action provides a generating
principle for twistor diagrams.  Here it is simplest, however, to
convert the diagram formalism into a set of position space and
momentum space Feynman rules.  We first consider the 3-vertex.  This
arises from the $a^3$ term in the Chern-Simons form and so the
corresponding 3-vertex will arise if we put three on-shell linearized
fields $a_1$, $a_2$ and $a_3$ into the formula
$$
V(a_1,a_2,a_3)=\int \tr\, (a_1\wedge a_2\wedge
a_3)\wedge\delta^{\prime\prime\prime}(Z\cdot W)D^3Z\wedge 
D^3W\, .
$$ In order to obtain a non-trivial result, we cannot have that all of
the $a$s are functions of $Z$ (or all functions of $W$) as the forms
will wedge to give zero.  Thus we can take without loss of generality
either $a_1=a_1(Z)$, $a_2=a_2(Z)$ and $a_3=a_3(W)$ to give the $--+$
vertex or $a_1=a_1(W)$, $a_2=a_2(W)$ and $a_3=a_3(Z)$ to give the
$++-$ vertex.  We focus on the $--+$ case as clearly the $++-$ case
works similarly.  We first partially integrate using the twistor
transform 
$$
g_3(Z)=\int_{\PT\times^\E\PT|_{Z=\mbox{const.}}}
a_3(W)\wedge\delta^{\prime\prime\prime}(Z\cdot W) D^3W 
$$
as above.
 This then gives the formula for the vertex as
$$
V(a_1,a_2,a_3)
=\int_\PT \tr (a_1\wedge a_2\wedge g_3)\wedge D^3Z\, .
$$
To evaluate this, we note first the standard integral formula for a
self-dual Maxwell field $G_3$ in terms of a 
homogeneity degree $-4$ cohomology class $g_3$ is
$$
G_3=G_{3A'B'}(x)\d^2x^{A'B'}=\int_{p^{-1}(x)}g_3\wedge D^3 Z\, .
$$
Secondly, the Woodhouse (1985) explicit
representatives for a homogeneity degree $0$ class $a_1(Z)$ is simply
the $(0,1)$ part of $p^*A_1$ where $A_1$ is the corresponding
potential 1-form on $\E$.  But $a_1\wedge D^3Z= p^*A_1\wedge
D^3Z$ because wedging with $D^3Z$ projects out the holomorphic part
of a $1$-form.  Thus we can write 
$$
V(a_1,a_2,a_3)=\int_\PT \tr (A_1\wedge A_2\wedge g_3)\wedge D^3Z
=\int_\E \tr \, (A_1\wedge A_2\wedge G_3)\, .
$$
We note that this space time representative is precisely what we would
have obtained from the space-time action (\ref{spacetimeaction}).

This formula can now be evaluated on momentum eigenstates 
$$
A_1=\e^{ip_1\cdot x}p_{1A}\epsilon_{1A'}\d x^{AA'}\, , \quad
A_2=\e^{ip_2\cdot x}p_{2A}\epsilon_{2A'}\d x^{AA'}\, , \quad
G_3=\e^{ip_3\cdot x}\tilde p_{3A'}\tilde p{3B'}\d^2 x^{A'B'}\, , 
$$
where $p_1^{AA'}=p^A_1\tilde p^{A'}_1$, $\tilde
p^{A'}_1\epsilon_{1A'}=1$  etc..  We find, after some manipulations
the standard result
$$
V(a_1,a_2,a_3)=\delta^4(p_1+p_2+p_3)\frac{\langle p_{1} \cdot p_2\rangle^4}
{\langle p_{1}\cdot p_2\rangle\langle p_{2}
  \cdot p_3\rangle \langle p_{3}\cdot p_1\rangle}    
$$ where $\langle p_1\cdot p_2\rangle=p_{1A}p_2^A$ etc., the
(degenerate) MHV formula for the $--+$ vertex.  In order to obtain
this result, we need to use the fact for 3 complex null vectors, the
equation 
$p_1+p_2+p_3=0$ implies that either all the self-dual, or all the
anti-self-dual spinor constituents of the momenta
are proportional. That leads to relations that allow
one to eliminate the polarization spinors $\epsilon_1$ and
$\epsilon_2$ leaving the desired formula.

The propagator can also be represented on space-time following the
calculation in \S\ref{actionequ}.  This gives the representation for
the propagator as the two-point
function 
$$
\frac{s_{AA'}s_{BB'}\d^2x^{A'B'}\wedge\d^2y^{AB}}{(s\cdot s)^3}\, .
$$
This is a non-standard expression for the photon propagator because it
is appearing in the
field representation (i.e., on the 2-forms rather than the potential)
and so it is in fact the second derivative of the standard photon propagator.
This is related to the fact that the currents are also being
represented as 2-forms, i.e., as potentials for their usual 3-form
representation. 
On momentum space this propagator becomes $p_{A(A'}p_{B')B}/p\cdot p$.

\section{Conclusions}
We see that the propagators and vertices are identical to
those of the space-time Lagrangian (\ref{spacetimelag}) and the
computation of scattering amplitudes from these two theories should
therefore be the same at tree level.  However, more work is required
to test the equivalence or otherwise at the level of loops.  The
generation of all amplitudes from trivalent vertices is suggestive of
the BCFW relations.

The Chern-Simons theory is clearly suggestive of a B-model
twistor-string theory in ambitwistor space in which a D7-brane is
wrapped on $\A_\E$ (with $\bar\xi=0=\bar\eta$); this would, however,
be a non-standard construction in string theory as in the B-model
branes are usually wrapped on holomorphic cycles. We intend to return
to the construction of an ambitwistor-string theory in a subsequent
paper.

Clearly there is a good prospect of providing a firm basis to twistor
diagram theory.  The propagator and vertices are clearly those of
twistor diagrams, but there are a number of differences as well.  For
example, in the twistor diagram approach
both $\delta^{\prime\prime\prime}(Z\cdot W)$ and $(Z\cdot W)^{-4}$ are
represented by the latter, but distinguished by the choice of contour.
A more fundamental difference is that vertices have more
twistor-functions ending on a twistor than are usually allowed.  This
is perhaps the most basic expression of infrared divergences.  There
is clearly much useful work to be done to make contact with the work
of Hodges (2005) on the twistor diagrams for gauge theory.

\section*{References}

\smallskip

\noindent
Aganagic, M., and Vafa, C. (2004)  Mirror symmetry and supermanifolds,
hep-th/0403192.

\smallskip

\noindent
Atiyah, M.F., Hitchin, N.J.\ and Singer, I.M.\ (1978) Self-duality in
four dimensional Riemannian geometry, {\it  Proc. Roy Soc. Lond.},
{\bf A 362}, 425-61.

\smallskip

\noindent
Baston, R.J. \& Mason, L.J. (1987) Conformal gravity and spaces of
complex null geodesics, Class. Quant. Grav., {\bf 4}.No.\ 4, 815-26.

\smallskip

\noindent
Britto, R., Cachazo, F., and Feng, B. (2005) New recursion relations
for tree diagrams of gluons, Nucl.\ Phys., {\bf B715},
hep-th/0413208.  

\smallskip

\noindent
Britto, R., Cachazo, F., Feng, B. and Witten, E. (2005) Direct proof
of tree level recursion relations in Yang-Mills theory,
hep-th/0501052.

\smallskip

\noindent
Cachazo, F., and Svrcek, P. (2005) Lectures on twistor strings and
perturbative Yang-Mills theory, arXiv:hep-th/0504194.


\smallskip

\noindent
Chalmers, G., and Siegel, W. (1996) The self-dual sector of QCD amplitudes,
Phys. Rev. D54, 7628-33.
arXiv:hep-th/9606061.

\smallskip

\noindent
Eastwood, M.G., and LeBrun, C (1986) Thickening and supersymmetric
extensions of complex manifolds, {\em Amer.\ J.\ Math.}, {\bf 108},
no.\ 5, 1177--1192.


\smallskip

\noindent
Ferber, A. (1978) Super-twistors and conformal supersymmetry,
Nucl. Phys., {\bf B132}.


\smallskip

\noindent
Harnad, J., Hurtubise, J., Legare, M., and Schnider, S. (1985)
Constraint equations and field equations in supersymmetric $N=3$
Yang-Mills theory, Nucl.\ Phys., {\bf 256}

\smallskip

\noindent
Hodges, A. (2005) Twistor diagram recursion for all gauge theory
amplitudes, hep-th/0503060.

\smallskip

\noindent
Isenberg, J., Yasskin, P., and Green, P. (1978) Non self-dual gauge
fields, Phys. Lett., {\bf
  78B}, 462-4.
See also: Isenberg, J., and Yasskin, P.\ (1979) in {\em Complex Manifold
  Techniques in Theoretical Physics}, Eds Lerner \& Sommers, Pitman.

\smallskip

\noindent
Manin, Y.I. (1988) Gauge field theory and complex geometry, GMW 289,
Springer.  

\smallskip

\noindent
Mason, L.J.\ (2005) Twistor actions for non-self-dual fields, a
new foundation for twistor-string theory, JHEP {\bf 10}, 009, hep-th/0507269.

\smallskip

\noindent
Mason, L.J., Singer, M.A., and Woodhouse, N.M.J.\ (2002)
Tau functions, twistor theory and quantum field theory, Comm.\ Math.\
Phys.\, {\bf 230}, no.\ 3, 389-420, arXiv: math-ph/0105038.



\smallskip

\noindent
Neitzke, A., and Vafa, C. (2004) $N=2$ strings and the twistorial
Calabi-Yau, hep-th/0402128.


\smallskip

\noindent
Prem Kumar, S., and Policastro, G. (2004) Strings in twistor
superspace and Mirror symmetry, hep-th/0405236.


\smallskip

\noindent
Witten, E. (1978) An interpretation of classical Yang-Mills theory,
Phys. Lett., {\bf 77B}, 394-8.

\smallskip

\noindent
Witten, E. (2004)  Perturbative gauge theory as a string theory in
twistor space, {\it Comm. Math. Phys.}, {\bf 252}, p189,
arXiv:hep-th/0312171. 

\smallskip

\noindent
Woodhouse, N.M.J (1985) Real methods in twistor theory, Class.\
Quant.\ Grav.\, {\bf 2}, 257-91.

\end{document}